# Exploring the Psychometric Validity of AI-Generated Student Responses: A Study on Virtual Personas' Learning Motivation

Huanxiao Wang
*Division of Human Development & Quant Methods*
*University of Pennsylvania*
*Philadelphia, PA, USA*
Contact: huanxiao0210@gmail.com

**Abstract**

This study explores whether large language models (LLMs) can simulate valid student responses for educational measurement. Using GPT-4o, 2000 virtual student personas were generated. Each persona completed the Academic Motivation Scale (AMS). Factor analyses(EFA and CFA) and clustering showed GPT-4o reproduced the AMS structure and distinct motivational subgroups.

## 1 Introduction

In psychometric research, collecting real human responses is essential for developing and validating psychological scales. Traditional item development requires large datasets to estimate item parameters, assess factor structures, and revise items based on empirical evidence. However, collecting human data is time-consuming, costly, and sometimes limited by ethical or logistical concerns.

With the rise of generative AI models, some researchers have explored using Large Language Models (LLMs) to simulate response data for psychological research. These models can generate both participant profiles (personas) and their responses to standardized instruments. For example, De Winter et al. (2024) used ChatGPT-4 to generate 2000 virtual personas who completed multiple personality tests. Other studies have applied similar methods in areas such as personality assessment (Argyle et al., 2023), item generation (Bhandari et al., 2024), and clinical diagnosis (Cook et al., 2024). These early findings suggest that LLMs may reproduce some aspects of human psychological variability, but the extent of their validity remains an open question. Liu et al. (2025) found that LLM-generated responses cannot fully substitute human respondents in all aspects of item-level psychometric performance.

In the field of educational psychology, relatively few studies have examined the use of generative AI models to simulate student motivation data. Learning motivation is commonly assessed through self-report instruments such as the Academic Motivation Scale (AMS; Vallerand et al., 1992) and the Motivated Strategies for Learning Questionnaire (MSLQ; Pintrich et al., 1991). Among these, AMS is widely used to measure motivation based on Self-Determination Theory (SDT; Deci & Ryan, 2000), covering intrinsic motivation, extrinsic regulation, and amotivation dimensions. The AMS has been validated in many populations and contexts (Bacanlı & Sahinkaya, 2011; Barkoukis et al., 2008; Fairchild et al., 2004; Guay et al., 2014; Stover et al., 2012; Utvær & Haugan, 2016), but it remains unclear whether AI-generated data can replicate its established factor structure.

This study investigates whether GPT-4o can simulate plausible student responses on the AMS, and whether the generated data exhibit acceptable psychometric properties. Specifically, this study asks:

**Q1:** Can GPT-generated responses reproduce the expected 7-factor structure of AMS, including IMTK, IMTA, IMES, EMID, EMIN, EMEX, and AMOT subscales?

**Q2:** Can GPT-generated persona descriptions be clustered into meaningful subgroups, and do these subgroups show distinct response patterns across AMS subscales?

## 2 Related Works

### 2.1 Measuring Learning Motivation

Student motivation has been widely studied in educational psychology. Several validated scales are used to measure motivation constructs. The Academic Motivation Scale (AMS; Vallerand et al., 1992) is one of the most widely used instruments, grounded in Self-Determination Theory (SDT; Deci & Ryan, 2000). The AMS assesses seven subtypes of academic motivation, including intrinsic motivation to know (IMTK), intrinsic motivation to accomplish (IMTA), intrinsic motivation to experience stimulation (IMES), identified regulation (EMID), introjected regulation (EMIN), external regulation (EMEX), and amotivation (AMOT). This structure has been validated across different populations and cultural contexts (Bacanlı & Sahinkaya, 2011; Barkoukis et al., 2008; Fairchild et al., 2004; Guay et al., 2014; Stover et al., 2012; Utvær & Haugan, 2016).

However, even for well-established scales like AMS, researchers typically require large-scale human response data to examine factor structure, reliability, and construct validity. This process can be resource-intensive, and difficult to replicate across diverse samples.

### 2.2 AI-Generated Psychometric Data

With the development of generative AI models, researchers have begun to explore using large language models to simulate human response data. De Winter et al. (2024) demonstrated that ChatGPT-4 can generate thousands of virtual personas who complete various personality inventories. Similar methods have been applied to simulate item-level responses in personality assessment (Argyle et al., 2023), clinical measurement (Cook et al., 2024), and item generation for educational testing (Bhandari et al., 2024). These studies suggest that LLMs may capture certain aspects of psychological variability.

However, the validity of AI-generated psychometric data remains uncertain. Liu et al. (2025) found that while LLM-generated respondents may mimic some human response patterns, they cannot fully substitute for human data, particularly when analyzing fine-grained item functioning. Moreover, most prior studies have focused on personality or clinical scales. Applications of LLMs in simulating student motivation data remain scarce.

### 2.3 Research Gap

Although early work suggests LLMs have some capacity to generate psychologically meaningful data, few studies have systematically tested whether AI-generated responses can reproduce complex factor structures of educational motivation instruments like AMS. In addition, the combination of LLM-based persona generation and psychometric analysis has not been fully explored in this domain. This study addresses this gap by evaluating whether GPT-4o can simulate realistic AMS responses, and whether persona embeddings can reveal subgroup patterns consistent with motivation theory

## 3 Methods

### 3.1 Participants and Data Generation

In this study, no real human participants were recruited. Instead, all data were generated through simulated personas using the GPT-4o model. The generation process included two stages: persona creation and questionnaire response simulation.

### 3.2 Personas Generation

The design of the persona prompt was informed by prior work using ChatGPT to create fictional respondents for psychological surveys (De Winter et al., 2024). The process was not a strict iterative protocol, but the final version was tested informally to ensure that the personas were coherent and diverse.

The prompt structure also reflects the RISE framework of prompt engineering. It defined a clear *Role* for the model (generate fictional students), specified the *Input* (age, gender, and a short profile), outlined the *Steps* (produce three descriptive sentences), and set the *Expectations* (concise one-line outputs). This helped create consistent personas while still allowing variation across individuals.

First, 2000 virtual student personas were generated using GPT-4o (temperature = 1). Each persona included three elements: age (18-25), gender, and a short description (3 sentences) summarizing their academic personality, learning style, and motivational tendencies. The generation prompt was structured to ensure diversity across motivational profiles while maintaining coherence within each persona. The personas were returned as text files for further processing. The prompts used in this stage are shown below:

*Generate 20 fictional student personas. Each should include:*
*- Age (18–25)*
*- Gender*
*- A 3-sentence description of their academic personality, learning style, and motivation.*
*Each persona should be on one line, like:*
*0001. 20, Female - Loves collaborative learning; often uses concept maps to organize her thoughts; tends to get anxious during exams.*
*Only return the 20 personas, nothing else.*

To generate the full dataset, GPT-4o was called repeatedly in batches of 20 personas per request. In total, 100 batches were generated to produce 2000 unique persona descriptions.

It is important to clarify that the inclusion of "learning styles" in the persona descriptions does not reflect a theoretical endorsement of this concept. The idea that students learn best in their preferred style has been widely challenged in the literature (Nancekivell et al., 2019). In this study, learning style phrases were used only to enrich the variety and naturalness of the personas. They were not analyzed as variables and did not influence the psychometric results.

### 3.3 AMS Responses Generation

After generating the personas, each simulated student was asked to complete the Academic Motivation Scale (AMS), which consists of 28 items rated on a 7-point Likert scale (1 = Does not correspond at all, 7 = Corresponds exactly). For response generation, GPT-4o was instructed to simulate AMS item-level answers based on the persona descriptions. To minimize randomness in item response generation, the temperature was set to 0. The model returned raw item responses as a list of integers for each persona. Here is the prompt used in this stage.

*Imagine the following student: (personas),*
*This student is now responding to the Academic Motivation Scale (AMS).*
*There are 28 items, each rated from 1 (Does not correspond at all) to 7 (Corresponds exactly).*
*(28 full items),*
*Please return exactly 28 integers separated only by commas. No explanation, no labels. Just the numbers.*

The full item texts were embedded into the prompt to ensure that GPT-4o received the exact questionnaire content for each response simulation.

### 3.4 Psychometric Analysis (EFA and CFA)

Exploratory factor analysis (EFA) was conducted to examine whether the GPT-generated AMS responses reproduced the expected factor structure. All 28 AMS items were included in the analysis. The factor extraction was performed using principal axis factoring. The number of factors was determined based on parallel analysis and scree plot inspection. Promax rotation was applied to allow correlated factors. Factor loadings were evaluated to assess whether items loaded onto the intended subscales.

Confirmatory factor analysis (CFA) was used to test the fit of the established seven-factor structure of AMS. Each item was assigned to its corresponding subscale based on the original AMS theoretical model (Vallerand et al., 1992). The model included the following factors: IMTK, IMTA, IMES, EMID, EMIN, EMEX, and AMOT. CFA was conducted using maximum likelihood estimation. Model fit was evaluated using common fit indices: Comparative Fit Index (CFI), Root Mean Square Error of Approximation (RMSEA), and Standardized Root Mean Square Residual (SRMR). Factor loadings were examined to assess item performance within each factor.

### 3.5 Semantic Clustering of GPT-Generated Personas

In addition to analyzing AMS responses, persona descriptions generated by GPT-4o were analyzed to identify motivational subgroups. The persona texts were vectorized using GPT-4o embedding models (text-embedding-3-small). The resulting embeddings represented the semantic information contained in each persona description. K-means clustering was applied to the embeddings to partition the personas into three clusters (k = 3). The choice of three clusters was based on initial exploratory analysis and interpretability considerations.

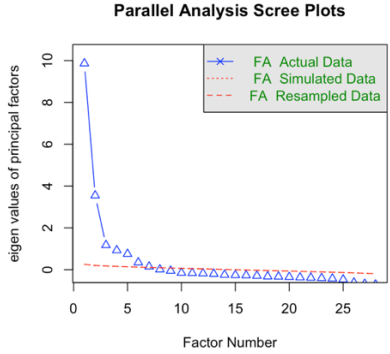

Figure 1: Parallel Analysis Scree Plot for Factor Extraction.

| Factor | Item | Standardized Loading |
|---|---|---|
| IMTK (Intrinsic Motivation - To Know) | AMS_Q2 | 0.805 |
| | AMS_Q9 | 0.938 |
| | AMS_Q16 | 0.953 |
| | AMS_Q23 | 0.857 |
| IMTA (Intrinsic Motivation - Toward Accomplishment) | AMS_Q6 | 0.791 |
| | AMS_Q13 | 0.582 |
| | AMS_Q20 | 0.818 |
| | AMS_Q27 | 0.778 |
| IMES (Intrinsic Motivation - Experience Stimulation) | AMS_Q4 | 0.417 |
| | AMS_Q11 | 0.926 |
| | AMS_Q18 | 0.980 |
| | AMS_Q25 | 0.899 |
| EMID (Extrinsic Motivation - Identified Regulation) | AMS_Q3 | 0.756 |
| | AMS_Q10 | 0.938 |
| | AMS_Q17 | 0.896 |
| | AMS_Q24 | 0.891 |
| EMIN (Extrinsic Motivation - Introjected Regulation) | AMS_Q7 | 0.852 |
| | AMS_Q14 | 0.814 |
| | AMS_Q21 | 0.851 |
| | AMS_Q28 | 0.807 |
| EMEX (Extrinsic Motivation - External Regulation) | AMS_Q1 | 0.758 |
| | AMS_Q8 | 0.891 |
| | AMS_Q15 | 0.863 |
| | AMS_Q22 | 0.978 |
| AMOT (Amotivation) | AMS_Q5 | 0.731 |
| | AMS_Q12 | 0.352 |
| | AMS_Q19 | 0.833 |
| | AMS_Q26 | 0.322 |

Table 1: Confirmatory Factor Analysis Results for GPT-Generated AMS Responses.

### 3.6 Subgroup Responses Analysis

After clustering, subscale scores for each of the seven AMS factors were calculated for each persona. Subgroup differences across clusters were analyzed to examine whether the clustering structure aligned with meaningful motivational patterns. Boxplots were used to visualize subscale score distributions across clusters. Non-parametric tests (e.g., Kruskal-Wallis tests) were performed to evaluate the statistical significance of subgroup differences on each AMS subscale.

All analyses were conducted in R (version 4.2.3; R Core Team, 2023) using RStudio (version 2025.05.1+513; Posit Software, PBC).

## 4 Results and Discussions

### 4.1 EFA and CFA

A parallel analysis was conducted to determine the appropriate number of factors. The analysis suggested that seven factors should be extracted, fully consistent with the original structure of the AMS. This result indicates that the GPT-generated responses preserved the intended dimensional structure of academic motivation as specified by Self-Determination Theory. Figure 1 shows parallel analysis scree plot for factor extraction.

The scree plot suggested a dominant first factor, followed by a sharp decline after the second factor. This result reflects a potential tendency of GPT-generated data to compress variance into fewer principal components, possibly due to the semantic coherence of AI-simulated responses.

A confirmatory factor analysis (CFA) was conducted to test the fit of the theoretical seven-factor structure of AMS. The CFA model included all seven subscales: IMTK, IMTA, IMES, EMID, EMIN, EMEX, and AMOT. The model demonstrated acceptable fit: CFI = 0.908, TLI = 0.894, RMSEA = 0.082, and SRMR = 0.065. These indices suggest that the GPT-generated responses generally reproduced the expected factor structure of AMS, although the fit was not perfect.

Standardized factor loadings were strong for most items, particularly for intrinsic and identified motivation subscales. For example, IMTK items loaded between 0.81 and 0.95, and EMID items loaded between 0.76 and 0.94. In contrast, several AMOT items displayed lower or unstable loadings (e.g., AMS_Q12 = 0.35; AMS_Q26 = 0.32). This pattern suggests that GPT-4o simulated positively valenced motivation dimensions more consistently than disengagement states such as amotivation.

The EFA and CFA results suggest that GPT-4o can partially reproduce the established factor structure of the AMS. While the model fit is not perfect, the seven-factor structure generally holds, particularly for intrinsic and identified motivation subscales. The relatively weaker performance on amotivation items may reflect GPT-4o's default bias toward goal-directed, coherent outputs, which may limit its ability to fully simulate psychological disengagement. Similar limitations have been noted in previous LLM-based simulation studies (Liu et al., 2025). Overall, these results provide preliminary evidence that LLM-generated response data may capture key aspects of psychological constructs but may require further refinement when modeling negative or conflictual motivational states.

### 4.2 Semantic Clustering of GPT-Generated Personas

To explore potential subgroups in the GPT-generated student personas, semantic clustering was conducted based on persona descriptions. Each persona text was vectorized using GPT-4o embeddings (text-embedding-3-small), and k-means clustering was applied. The number of clusters was set to k = 3 based on interpretability and preliminary exploration.

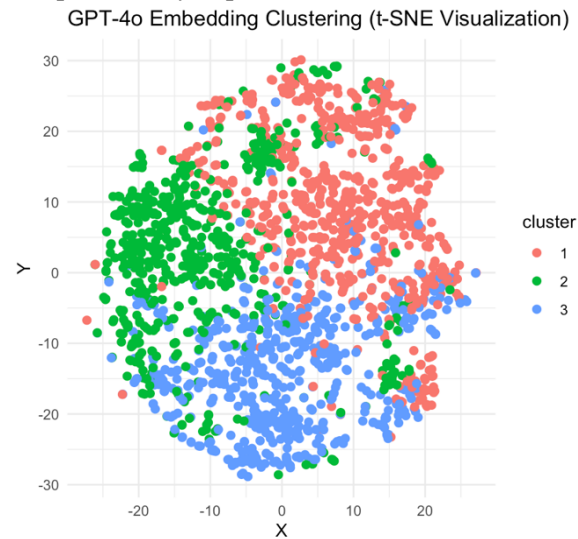

Figure 2: Semantic Clustering of GPT-Generated Personas Using t-SNE Visualization

A t-SNE visualization was generated to display the cluster separation in two dimensions. The clusters were well-separated, suggesting that GPT-generated persona descriptions contained distinct semantic features that could differentiate students into subgroups.

Subscale scores for the seven AMS factors were calculated for each persona. Boxplots were created to compare AMS subscale scores across clusters. Results showed that cluster membership was associated with different motivational profiles.

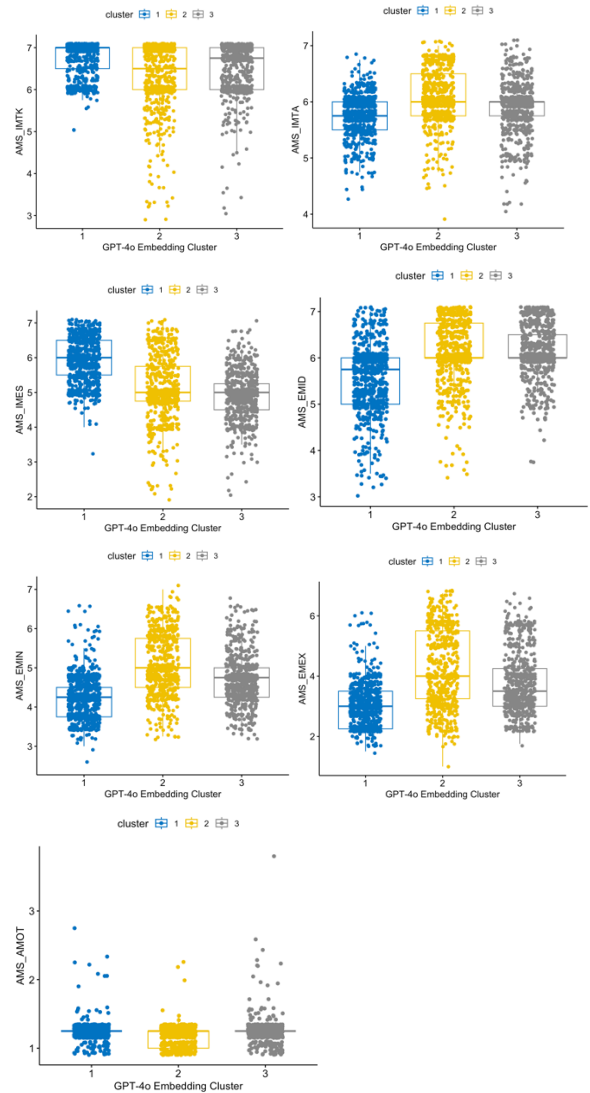

Figure 3: Subscale Score Differences Across GPT-4o Semantic Clusters

Cluster 1 displayed higher scores on intrinsic motivation subscales (IMTK, IMTA, IMES), suggesting students with strong intrinsic academic interests. Cluster 2 showed moderate levels of intrinsic and extrinsic motivation. Cluster 3 demonstrated slightly elevated external regulation (EMEX) and lower intrinsic motivation, suggesting more externally driven or performance-oriented students.

Kruskal-Wallis tests confirmed significant differences across clusters for most AMS subscales

(p < .001 for IMTK, IMTA, EMEX, AMOT; p < .01 for IMES, EMID, EMIN), indicating that semantic clustering based on persona descriptions corresponded meaningfully with simulated questionnaire responses.

The clustering results suggest that GPT-4o not only generated item-level responses consistent with AMS factor structure, but also produced semantically rich persona profiles that reflected distinct motivational orientations. Subgroups identified through semantic embeddings showed systematic differences across AMS subscales, supporting the convergent validity between generated persona characteristics and questionnaire outcomes.

These findings demonstrate that large language models may capture latent psychological patterns even before formal scale administration, purely based on persona-level text descriptions. This capability may have potential applications for early-phase scale development, where synthetic data may help evaluate item functioning across diverse hypothetical profiles prior to human data collection. Researchers can use this approach to screen for problematic items, evaluate whether expected factor structures emerge, and explore subgroup patterns across hypothetical profiles. Such applications may reduce costs, speed up validation cycles, and support scale adaptation in new contexts. Theoretically, this work also contributes to ongoing debates about the extent to which LLMs can approximate latent psychological constructs. It shows both the potential and the current limits of AI personas in capturing human-like motivational patterns.

However, it should be noted that GPT-generated clusters may reflect idealized or overly coherent motivational types, as the model tends to generate consistent and goal-oriented outputs. Additional validation with real human data is needed to fully assess the generalizability of these subgroup structures.

## 5 Limitations and Future Directions

While the present study demonstrates the promising potential of large language models (LLMs) like GPT-4o in generating psychologically plausible student response data, several limitations should be acknowledged.

First, although the generated responses reproduced the theoretical factor structure of AMS reasonably well, the confirmatory factor analysis still yielded only moderate model fit (e.g., RMSEA = 0.082). This suggests that LLM-generated data may not fully replicate the nuanced variance found in real human populations. In particular, the amotivation (AMOT) subscale consistently showed weaker or unstable loadings, which may reflect GPT-4o's inherent difficulty in simulating disengaged or conflicted psychological states. This aligns with prior observations that LLMs tend to default toward coherent, goal-oriented, and positively valenced outputs (e.g., Liu et al., 2025).

Second, the current study focused on only one questionnaire (AMS) and one LLM model (GPT-4o). The generalizability of these findings to other constructs, instruments, or LLM architectures remains unclear. Expanding this approach to include additional validated scales (e.g., MSLQ, AEQ) and cross-model comparisons would help clarify whether the observed psychometric patterns are robust across different psychological domains.

Third, the clustering analysis relied solely on text embeddings of GPT-generated persona descriptions. While meaningful subgroups were identified, these clusters are not directly validated against real-world student samples. Future work should compare LLM-generated subgroup structures to empirical cluster solutions obtained from human data to assess alignment and potential biases.

Finally, this study examined LLM-generated responses under controlled prompting conditions, using fixed temperature settings and instruction formats. Prompt engineering decisions likely play a crucial role in shaping response variability and latent structure reproduction. Future research should systematically investigate how prompt design, randomness parameters, and persona context framing influence the psychometric properties of generated data.

Another point worth noting is that the use of learning style descriptors in the persona prompts should not be read as a validation of the learning styles hypothesis. Like we mentioned before, the concept has been widely debated and is not supported by strong empirical evidence (Nancekivell et al., 2019). Here it served only as a descriptive element to make the personas sound more realistic, and it had no bearing on the psychometric findings.

Despite these limitations, this study offers a novel empirical demonstration of how generative AI can contribute to early-stage scale development and psychometric exploration. As LLM capabilities continue to evolve, careful validation studies combining both real and simulated data will be critical for evaluating the responsible integration of AI tools in psychological measurement. The findings have both theoretical and practical implications. Theoretically, they suggest that large language models can reproduce complex motivational structures like those in AMS, although with biases toward positive and coherent states. This adds to current discussions in psychometrics about whether AI can model latent constructs. Practically, the study points to the value of AI personas as a tool for instrument testing and development. With further refinement, such simulations may help researchers reduce the cost and time of scale validation, while also expanding opportunities to explore item functioning across diverse cultural or contextual settings.

## Acknowledgments

The author declares no conflict of interest. The datasets generated during this study are available from the corresponding author upon reasonable request. This paper was completed as part of an independent study during the author's master's program; all views and any errors are the author's own and do not represent the University of Pennsylvania. The author also used OpenAI's ChatGPT to assist in polishing the language of the manuscript.